\documentclass[preprint,11pt]{aastex}

\usepackage{hyperref}

\usepackage{graphicx}
\usepackage{natbib}
\usepackage{wrapfig}

\newcommand{\be}{\begin{equation}}
\newcommand{\ee}{\end{equation}}
\newcommand{\nn}{\mbox{} \nonumber \\ \mbox{}}
\newcommand{\ba}{\begin{eqnarray}}
\newcommand{\ea}{\end{eqnarray}}
\newcommand{\om}{\omega}

\newcommand{\E}{{\bf E}}
\newcommand{\B}{{\bf B}}

\newcommand{\Bf}{{magnetic field}}

\newcommand{\NS}{neutron star}
\newcommand{\ms}{magnetosphere}
\newcommand{\NSs}{{neutron stars}}
\newcommand{\mss}{magnetospheres}

\newcommand{\BH}{{black hole}}
\newcommand{\BHs}{{black holes}}
\newcommand{\EM}{{electromagnetic}}
\newcommand{\Sch}{Schwarzschild}
\newcommand{\LC}{light cylinder}
\newcommand\eg{\it{{e.g.,}}}

\newcommand\etal{\it{et al.}}
\newcommand\lo{\mathrel{\raise.3ex\hbox{$<$}\mkern-14mu\lower0.6ex\hbox{$\sim$}}}
\newcommand\go{\mathrel{\raise.3ex\hbox{$>$}\mkern-14mu\lower0.6ex\hbox{$\sim$}}}

\usepackage{amsmath}
\usepackage[normalem]{ulem}
\usepackage{color}

\begin{document}


\title{Multi-Messenger  Windows on the Universe: detecting precursor emission to compacts' mergers}
\author{Maxim Lyutikov\\
Department of Physics, Purdue University, \\
 525 Northwestern Avenue,
West Lafayette, IN
47907-2036 }

\begin{abstract}

We provide an overview of various mechanisms, and corresponding powers, of precursor emission to compacts' mergers to be detected by LIGO-Virgo-KAGRA  (LVK)  collaboration. Expected peak powers, $\leq  10^{43}$  erg s$^{-1}$,  are not sufficiently high to be detected by all-sky high-energy satellites (unless beamed). The best chance is the detection of possible coherent radio emission, producing observable signals up to $\sim$ Jansky of flux density. Low-frequency phased array telescopes like LOFAR, the  MWA  and DSA-2000 are best suited  due to their large instantaneous sky coverage. Time-wise, in addition to LIGO  early warning alerts up to a minute before the merger, the dispersive delay at lower frequencies of $\sim$ 300 MHz can be of the order of minutes.  Optical detections are the most challenging. 
\end{abstract}

\tableofcontents

LIGO early warning alerts \citep{2020ApJ...905L..25S},  \url{https://emfollow.docs.ligo.org/userguide/early_warning.html}  are  expected to provide localization to $\sim 300-1000$ square degrees up to a minute before merger.  This opens a possibility of observing precursor \EM\ emission to gravitational waves sources.   Here we discuss the expected powers, modulation, and corresponding observing strategies. 

Compacts' mergers  (double {\NS}s, DNS, \NS-\BH, and double {\BH}s) are expected to occur in old systems, with not much surrounding material. 
Thus,  little  pre-merger accretion power  is expected \citep{2018PhRvD..98h1501F,2020PhR...886....1N}.  Tidal disruption during active stages of  mergers 
\footnote{For BH-NS mergers, a BH of mass $\geq 5 M_\odot$ will not tidally disrupt the NS.} will produce optically thick material that will likely  blanket instantaneous radiation, see though \S \ref{hair1}. In addition, prospective sites of   coherent radio emission - the likeliest detectable candidate, see below - will likely be polluted and  radio emission switched off by the abundant tidally ejected material.

Observable  radiation, if any, may still come from the  \EM\ interaction of merging stars. The underlying mechanism is the variant of unipolar induction operating in highly magnetized relativistic plasma self-produced  by the merging companions.
The basic estimate of EM power expected from a relativistic unipolar inductor \citep[{\eg}][]{2002luml.conf..381B} is 
\be
L\sim  \frac{c}{4\pi} (\Delta \Phi)^2
\label{L} 
\ee
where 
\be 
 (\Delta \Phi) \sim \beta D B
 \ee
 is the EMF drop for a conductor of size $D$ moving in \Bf\ $B$ with velocity $v = \beta c$.  The factor $4\pi/c= 377 $ Ohm is sometimes called  the impedance of free space.
 
 For rotating neutron  stars, the potential drop $\Delta \Phi$ is over the \LC: $\Delta \Phi \sim B_{NS} R_{NS}^3 (\Omega_{NS}/c)^2$ \citep{goldreich_julian_69}.
 For a  linearly moving \NS\ or \BH\  the potential drop $\Delta \Phi$ is over the linear dimension of a conducting body 
 \citep{1969ApJ...156...59G,2011PhRvD..83f4001L}.
 \footnote{
An alternative  popular scaling of \EM\  power is 
 $
 L\sim  B^2 D^2 V
 $   \citep[][]{2007P&SS...55..598Z}. It is larger than (\ref{L}) by a factor $\sim c/V \gg 1 $. In this case a star (or  a planet) is assumed  to  ``eat as Pac-man" all the energy of the \Bf\   incoming within its cross-section (or, even higher, within the cross-section of the magnetopause). Such scalings highly   over-estimates the  dissipated \EM\ power (see, {\eg}   \cite{1969ApJ...156...59G}  and Eq. (28) of 
 \cite{1980JGR....85.1171N}.
}

 Next,  we discuss various applications on the   relation (\ref{L}) to LVK and LISA sources: Keplerian motion of merging objects, \NSs\ and \BHs, as well as spinning magnetized \NSs\ and \BHs.
 
 \section{Magnetospheric interaction of merging \NSs}
 \subsection{Estimates of the average power} 
 
\begin{figure}[h!]
 \includegraphics[width=.9\linewidth]{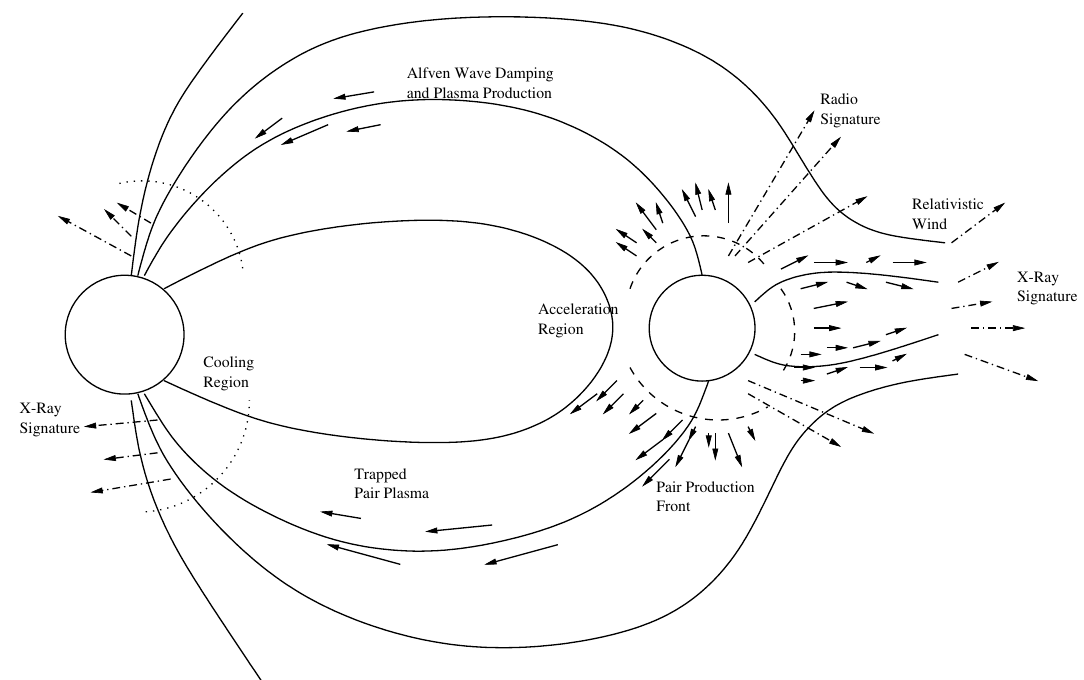}
\caption{Schematic version of the energy extraction process. The motion of the companion
through the magnetar field induces a plasma flow from the companion into the magnetosphere.
The pressure of this flow will drive a relativistic wind in those regions where the
flow moves into a regime of  a weaker field, while the plasma remains trapped in the case when it
flows into a stronger field regime. The hot pair plasma will ablate some baryons off the surface
of the neutron star, providing a baryon-loaded sheath that  regulates the cooling of the
trapped plasma \citep[after][]{2001MNRAS.322..695H}.}
\label{Drawing}
\end{figure}

 The first basic application of the  relation (\ref{L}) is to a merger of  an unmagnetized \NS\ (so, $D\sim R_{NS}$) moving with Keplerian velocity  through the \mss\ of the companion \citep{2001MNRAS.322..695H,2012ApJ...757L...3L}, Fig. \ref{Drawing}
 \be
 L_{1NS} \sim \frac{{B_{\rm NS}}^2 G M {R_{\rm NS}}}{c} \left( \frac{R_{NS}} {r} \right) ^7 = 
 \left\{
 \begin{array}{c}
 5 \times 10^{44} \, {\rm erg\,s}^{-1}  \left( \frac{R_{NS}} {r} \right) ^7
 \\
 9 \times 10^{37} \, {\rm erg\,s}^{-1} \left( \frac{-t}{ {\rm sec} }  \right)^{-7/4},
 \end{array}
 \right.
 \label{L1} 
 \ee
 where we used   time to merger  $t$ 
\begin{equation}
- t= \frac{5}{256} \frac{c^5}{G^3} \frac{r^4}{M_1 M_2 (M_1+M_2)}
\label{tm}
 \end{equation}
 is measured in seconds in Eq.  (\ref{L1}). (For numerical estimates we assume surface fields of $B_{\rm NS} =10^{12}$ G.
Due to tidal distortions, the minimal time in (\ref{tm}) is  $ \tau_0 \sim10^{-3} $ sec (this is the peak of LVK), giving the peak power of 
 \be
 L_{peak}  \sim 10^{43} \, {\rm erg\,s}^{-1} ,
 \label{Lpeak}
 \ee
 see  \citep{2001MNRAS.322..695H,2023ApJ...956L..33M}. 

Though the  relation (\ref{L})  appeals to induction  - a non-dissipative process - in reality,  the  inductively created electric fields can have a component along the total magnetic field (gaps) and/or the electric field may exceed the value of the local magnetic field, see below.

Relations (\ref{L}-\ref{Lpeak})   are important basic estimates. The peak power - emitted within a millisecond, is not large, making it challenging to detect. 

Another  important point about detectability  is the possible  orbital modulation of the emission.
If only one star is magnetized  the emission is likely to be produced along the direction of the \Bf\ at the location of the secondary; then, if the magnetic axis is misaligned with the orbital spin, this direction is modulated on the orbital period.

 If both stars are magnetized, the effective size of ``frictioning" \mss\ is the orbital separation, $D\sim r$. In this case,  the power of $R_{NS}/r$ in (\ref{L1})  is  $(R_{NS}/r)^5$ instead of power $7$ \citep{2019MNRAS.483.2766L,2022MNRAS.515.2710M}. So, larger ramp-up power, approximately the same peak power. 
 
\begin{figure}[h!]
 \centering 
\includegraphics[width=.6\columnwidth]{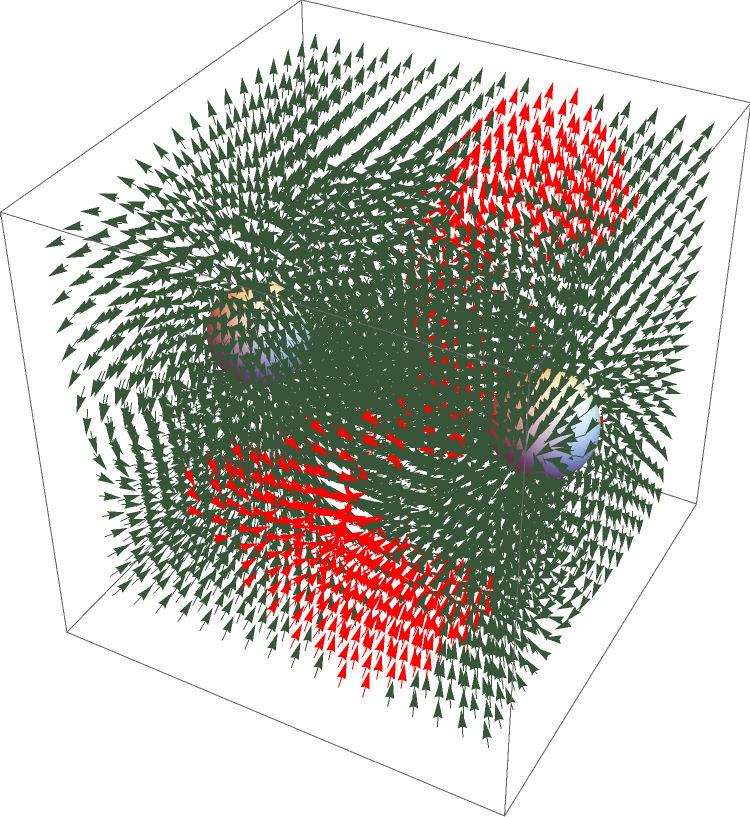} 
\caption{ 3D rendering of the common \ms\ of the interacting \NSs, orthogonal dipoles. Highlighted in red are regions with high parallel electric field \protect\citep{2019MNRAS.483.2766L}. }
 \label{Eparao}
\end{figure}

 For the double-magnetized case,  the structure of the common \ms\ of the non-rotating  \NSs\  is complicated, with gaps, but no $E>B$ regions, Fig. \ref{Eparao}. There is strong orbital variations for the case of misaligned magnetic moments. 
 
\subsection{The  merger of two NSs is   necessarily electromagnetically   dissipative}

\begin{figure}[h!]
 \includegraphics[width=.45\linewidth]{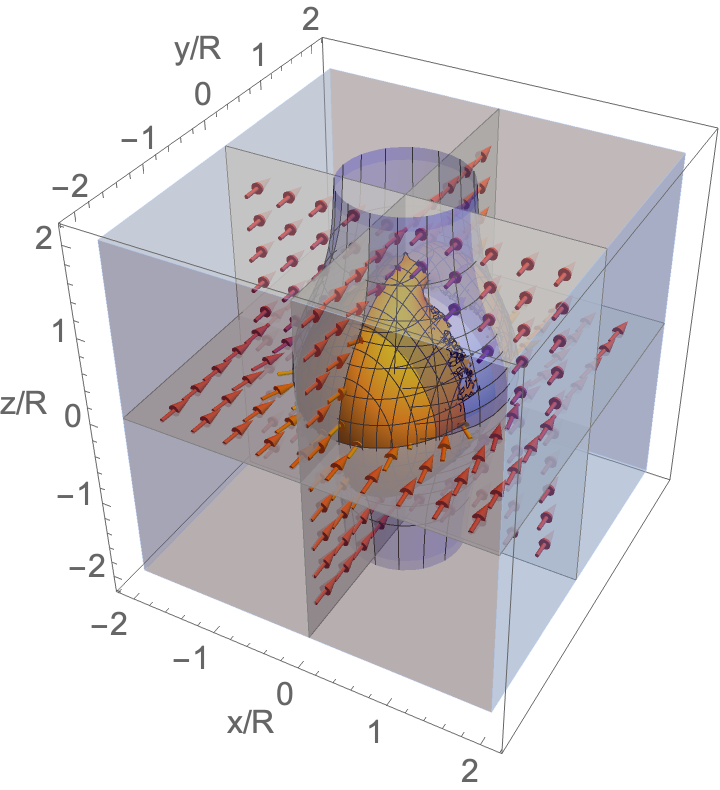}
\includegraphics[width=.54\linewidth]{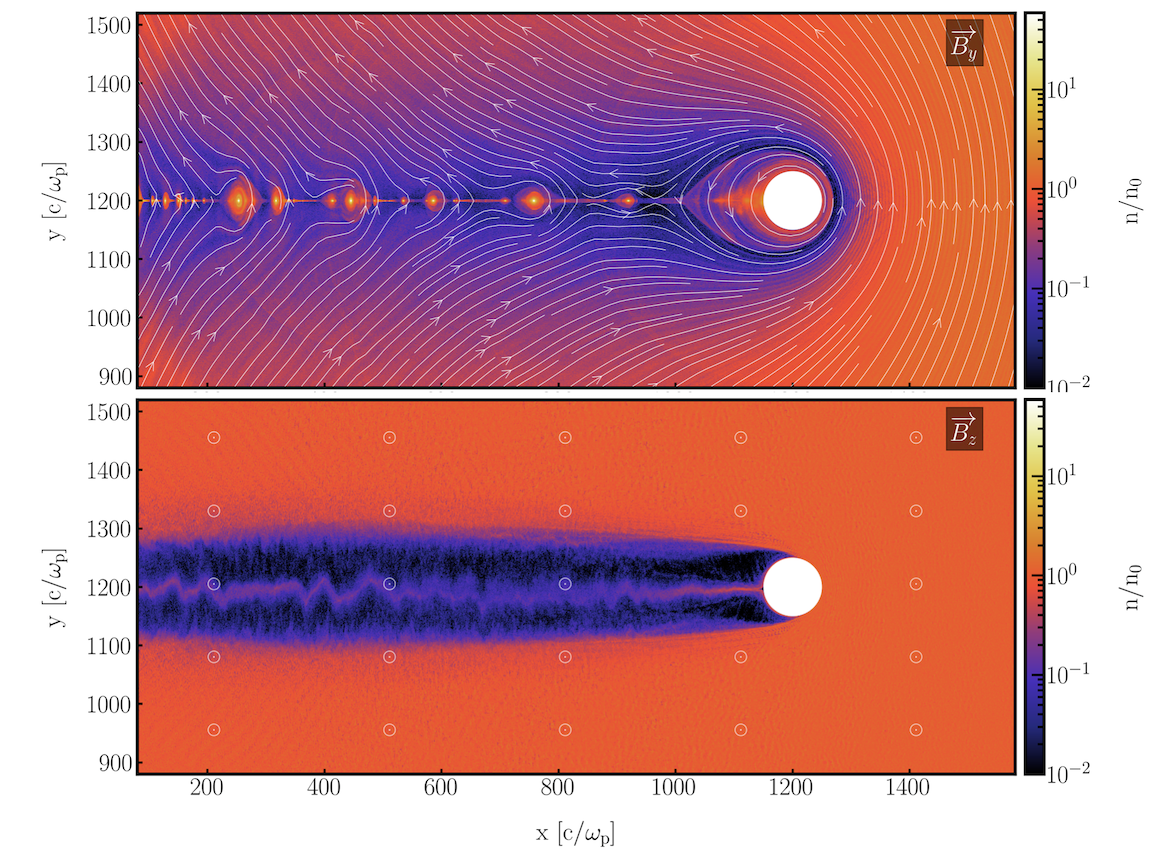}
\caption{ Left panel: 3D view of first order electric field near the \NS\ \protect\citep{2023PhRvE.107b5205L}. The central sphere is the \NS. The blue surface is the magnetic field flux surface (magnetic field lines lie on the surface pointing in the $z$ direction. Arrows are electric fields sliced at $x=0, y=0, z=0$. In the frame of the \NS\ plasma is moving in the $+x$ direction. Bounded ear-like surfaces are regions where $\beta_{EM} $ becomes larger than 1. Right Panel:  2D PIC simulations of relativistic magnetic draping, different orientations of the magnetic field (in the plane and out of the plane). Plotted is plasma density. Dissipative regions downstream, due to magnetic draping, are clearly seen (Cortes+, in prep.)}
\label{densbfieldlines1}
\end{figure}

The above discussion implicitly assumes that  the emitted  \EM\ energy comes from the kinetic energy of Keplerian motion, but it  does not address the question why and how the dissipation takes place. It turns out that 
{\it the electromagnetic interaction of merging neutron stars is necessarily dissipative}  due to the effect of electromagnetic draping - the creation of dissipative regions near the companion (unmagnetized) star \citep{2023PhRvE.107b5205L}. The draping effect is well-known in space and astrophysical plasmas \citep{2004AIPC..719..381C,2006MNRAS.373...73L,2008ApJ...677..993D}. In the conventional non-relativistic MHD limit the creation of the magnetized layer (for super-Alfvenic motion) does not lead to dissipation, only a breakdown of the weak-field approximation in the draping layer.

\subsection{Production of flares in DNS mergers} 
\label{2to1}

The above estimates of {\it average}  power  (\ref{L1}-\ref{Lpeak}) are relatively small. Importantly, those estimates assume continuous dissipation. Possible production of flares - whereby energy is slowly stored and then suddenly released - may produce (orbital-modulated) flares. Indeed, a magnetic connection between merging magnetized \NSs\ may lead to the production of a special configuration of interacting \mss\ of merging \NSs\ \citep{2021ApJ...923...13C}. 

In the  case of interacting  \mss, 
there are special configurations  when the stellar spins and the orbital motion nearly `compensate' each other, leading to  very {\it slow} overall winding of the coupled magnetic fields; slowly winding configurations allow   gradual accumulation of magnetic energy,  that is eventually released in a flare when the instability threshold is reached. This  slow winding can be  global and/or local.   The conditions  for the storage of  magnetic energy sometimes become ambiguous near the 
topological  bifurcation points;  in certain cases,  they also depend on the relative  phases of the spin and orbital motions.    In the case of merging  magnetized neutron stars,      if one of the stars is a millisecond pulsar, spinning at  $\sim$ 10 msec, the global  resonance $\omega_1+\omega_2=  2 \Omega$ (spin-plus beat is two times  the orbital period)    occurs approximately a second   before the merger; the total energy of the flare can be as large as $10\%$ of the total magnetic energy,  producing bursts of luminosity $\sim    10^{44}$ erg s$^{-1}$.   Higher order local resonances may have similar powers, since the amount of involved magnetic flux tubes may be comparable to the total connected flux.

 The 2:1 resonance  occurs at 
 time
 \begin{equation}
t_{1-2}=(-t_m)=  10^{-3} \frac{ c^5 (M_1+M_2)^{1/3}}{G^{5/3} M_1 M_2 \omega^{8/3}}= 5\times 10^{5} P_s^{8/3} \, sec
,\end{equation}
where $P_s = 2\pi/\omega$ is the sum-beat (addition)   period of the NS spin's, $M_1$ and $M_2$ are masses of neutron stars (assumed to be equal to $1.4M_\odot$. 
Thus, the most interesting case is one of the neutron star is a recycled millisecond pulsar. For example, if $P_s = 10 $ msec, then $(-t_m)=2$ sec. 
At that moment stars are separated by 
\begin{equation}
r= ( G (M_1+M2))^{1/3} \Omega^{-2/3} \approx 10^7 {\rm cm},
\end{equation}
approximately ten times the radius.

We can also estimate how flaring will evolve with time. At exact resonance there is no shear, no flaring. As the orbit shrinks,  the system  gets out of the resonance, field lines are becoming twisted. Assuming that flares occur after a fixed twist angle $\sim 1$, the time between flares evaluates to
\begin{equation}
t_f= \frac{4 P}{3 \pi} \frac{t_{1-2}}{t-t_{1-2}}
\end{equation}

\begin{figure}[h!]
\centering
\includegraphics[width=.6\textwidth]{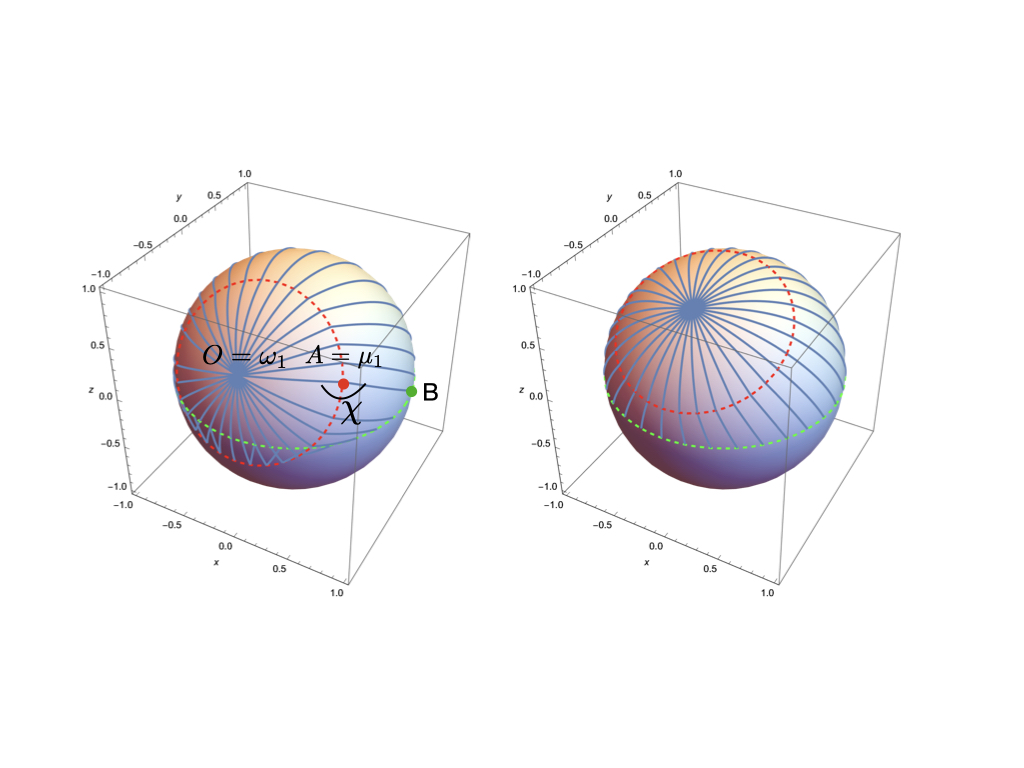}
\includegraphics[width=.39\textwidth]{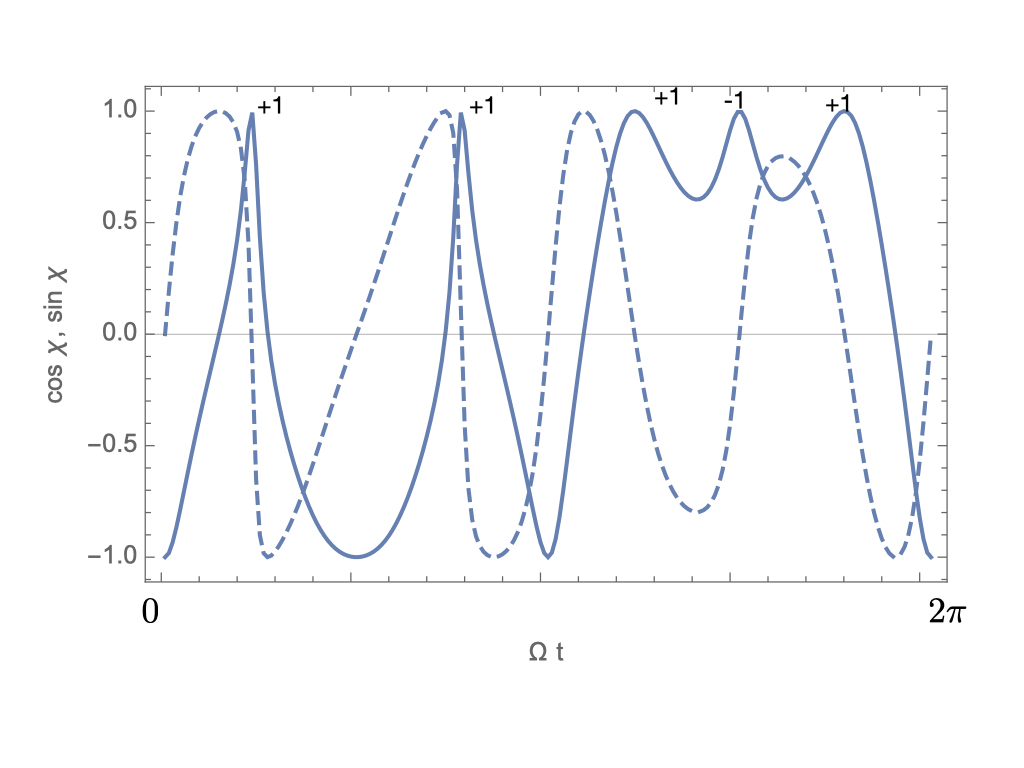}
\caption{ Left  panels: Twisting of coupled \mss. The sphere represents a space of the directions of the magnetic field. Red dashed circle is the trajectory of the direction of the  magnetic field at point $A$ (direction of magnetic moment of $1$); it rotates with the angular frequency $\omega_A=\omega_1$ around point $O$, the direction of the spin of the first star.
For both plots $\omega_1= \Omega$: one spin per one full orbital rotation. Left panel inclination $\theta_A =\pi/3$, right panel  $\theta_A =\pi/6$ (these are the angles between orbital momentum $\vec{\Omega}$ and the spin $\vec{\omega}_1$. Radius of the circle $\theta_{A,c} = \pi/4$ (this is the angle between the spin $\vec{\omega}_1$ and the magnetic field $\vec{\mu}_1$). Notice how arcs cross-over in the lower part in the left panel, when point $A$ is below the equator. 
Right panel: Counting the twist. Plotted is the angle $\chi$, spin of A - magnetic moment of A - direction to B, ($\cos \chi$ - solid line,  $\sin \chi$ - dashed line). One twist  is added to magnetic stripe connecting two stars when $\cos \chi=1 $ and  $\sin \chi$ changes from positive to negative; one rotation is subtracted when $\sin \chi$ changes from negative to positive. In this example $\omega_A = 5\Omega$,   $\theta_A =\pi/3$  and $\theta_{A,c} = \pi/4$ \protect\citep{2021ApJ...923...13C}. }
\label{Divingunder} 
\end{figure}

In addition to 2:1 global resonance, there are many other possible resonances of the type $m(\om_1+ \om _2) = n \Omega$.  Their occurrence depends on the particulars of geometry,   Fig. \ref{Divingunder}. Though in principle high-order resonances are possible ($m \gg n$), the more frequent $m \sim n$ impose a constraint: most powerful flares are expected close to the merger, when the orbital period is sub-second. To be in resonances, the spins should be similarly fast, but this may not be realized in the old \NSs. 
\citep[An example to the contrary is the Double pulsar][at the time of merger the companion PSR-A will be still in tens of milliseconds spins ]{2004Sci...303.1153L}

\begin{figure}[h!]
\centering
\includegraphics[width=.6\textwidth]{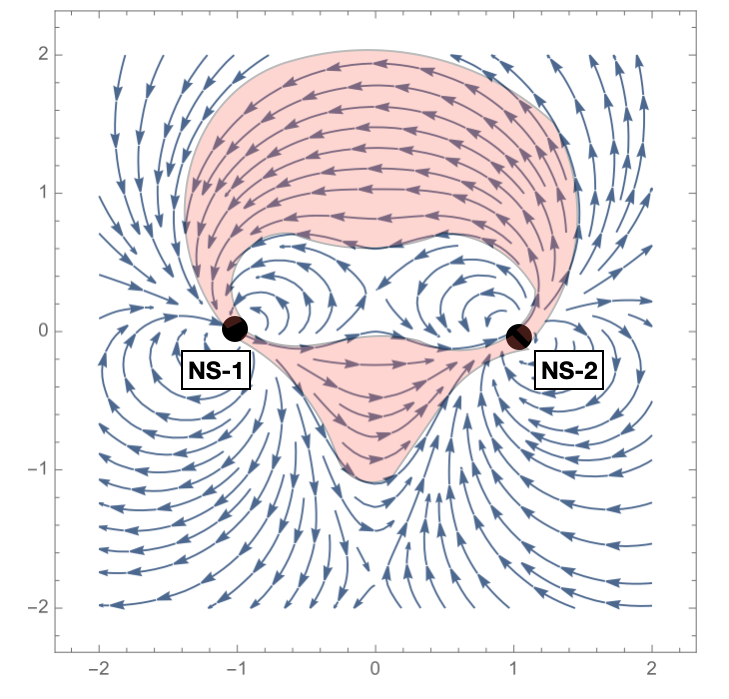}
\caption{Example of coupled {\NS}s \mss. The highlighted region contains field lines connecting two stars. Under certain conditions  \protect\citep{2021ApJ...923...13C} the \Bf\ lines in the coupled region are slowly twisted, reach an  instability point, and release magnetic energy on light travel time $\sim r/c$.}
\label{Divingunder1} 
\end{figure}

Energy-wise, higher order resonances  involves only those magnetic lines whose direction at the star surface lie in both the interacting magnetic polar cap  and in the resonant  region.  In  special  cases, even though the untwisting region is small, it may involve a large fraction of the  (also small) topologically  connected region.  Thus, 
a flare may release an amount of energy contained within the flux tube connecting two stars on reconnection time scale: the light time travel over the orbital separation,  $\sim r/c$, see Fig. \ref{Divingunder1}. The corresponding energies $E_f$  and luminosities  $L_f$ of the resulting flares estimate to  (upper limits)
\ba && 
E_f \sim B_{NS}^2 \frac{ R_{NS}^6}{r^3} = 10^{39} \left( \frac{-t}{ {\rm sec} }  \right)^{-3/4} \,  {\rm erg} \leq 2 \times  10^{41} {\rm erg}
\nn &&
L_t \sim \frac{E_f}{r/c} = 10^{42} \left( \frac{-t}{ {\rm sec} }  \right) ^{-1} {\rm erg\,  s} ^{-1} \leq  10^{45}  {\rm erg s} ^{-1} 
\ea
(upper limits are for $-t= 10^{-3} $ seconds). \citep[see also][]{2022MNRAS.515.2710M}.

  \subsection{Precursor  Cherenkov emission to BH-NS mergers}
  \label{Cherenkov}

BH-NS mergers can also produce low frequency (radio) Cherenkov emission \citep{2025PhRvD.111j3010K}.  
This is an unusual    effect at the intersection of gravity, electromagnetism and physics of continuous media:  Cherenkov emission by a  {\it classical  uncharged} \Sch\ \BH\ moving superluminally (with respect to the speed of light in the medium) through a dielectric in \Bf.

The governing equations (involving general relativity, electromagnetism,  and the physics of  continuous media)  have no external \EM\ source---it is the 
distortion of the initial  \EM\ fields by the gravity of the \BH\ that  plays the role of a superluminally moving  source.  The effect relies on nonzero values of both the magnetic field and the gravitational radius, as well as on the usual Cherenkov condition on the velocity, $v/c > 1/\sqrt{\epsilon}$.

Media with subluminal phase velocity are expected in astrophysical settings.  For example, strongly magnetized plasmas, with $\om_B \geq \om_p$ ($\om_p$ is plasma frequency, and $\om_B= e B/(m_e c) $ is cyclotron frequency)  support  subluminal normal modes \citep{1986ApJ...302..138B,1999JPlPh..62...65L}.
This is the plasma regime expected, {\eg}  around various types of \NSs. 

In the Cherenkov regime,  emission conditions involve the combination
\be
|\Lambda|  B_0 M \equiv 
\sqrt{ |\beta^2 \gamma^2 (\epsilon -1) -1|} B_0 (G M )
\label{factor}
\ee
It requires gravity, \Bf, and superluminal  motion.
The emission also  requires that the plasma motion is not force-free - there is a non-vanishing Lorentz force on plasma. Emisison is absent only in the case  if  the velocity of the medium is everywhere  along the {\it local}  \Bf\ ({\eg} there is emission if   a BH propgates along   the global \Bf,  but  plasma locally  sliding across the   field lines).

An order-of-magnitude estimate of Cherenkov power $L$ is 
\be
L \sim \beta_K B(r)^2 \left( \frac{2 G M_{BH}}{c^2} \right)^2 c   \approx  10^{38} \, {\rm erg\, s}^{-1} (-t)^{-13/8} 
\left( \frac{B_{NS}}{10^{12} {\rm G}}  \right) ^{2}\left( \frac{M_{BH}}{10 M_\odot}  \right) ^{-3/4}
\ee
 where $-t$ is time to merger in seconds (\ref{tm}),    The peak power, estimated at $-t \sim $ millisecond is $\sim $ few $10^{42}$ erg s$^{-1}$.   (Incidentally, the value of peak power comes close to the parameters of enigmatic Fast Radio Bursts \cite{2019ARA&A..57..417C}, but the expected event  rates  do not match.)
  
  On the one hand, the  peak luminosity is not high, similar to other mechanisms  based on unipolar inductor  scaling \citep{2001MNRAS.322..695H,2011PhRvD..83l4035L}. But it can be detectable in radio - the corresponding peak flux is in the kilo Janskys range if 
 coming from $\sim 100$ Mpc distance \cite{2024arXiv240216504L}.      
   An important constraint comes from the expansion in small mass of the BH, which translates to the requirement that the emitted wavelength should be larger than the \Sch\ radius. For stellar mass BHs the  \Sch\ radius is kilometers, this implies frequencies below the kilo-Hertz range. Depending on the parameters at the source, such low frequencies may  not be able to propagate  due to plasma cut-off at $\omega = \omega_p $.  The required  density is $\leq 1$ cm$^{-3}$. This is somewhat larger  than the typical density in the volume-dominant warm phase of ISM ($n _{ISM} \sim 10^{-2}- 10^{-1} $ cm$^{-3}$). 
   
   Primordial black holes \citep{2021RPPh...84k6902C} may  not suffer from such limitation -  their emitted frequency can be in the GHz range for masses $M_{BH} \sim 10^{-6} M_\odot$.  The corresponding power for a primordial black hole flying through the \ms\ of a neutron star is
     \be
  L_{\rm BH, 10^{-6} M_\odot, max} \approx 4 
  \times 10^{34} \, {\rm erg\, s}^{-1}
\label{LCH1}
 \ee

\subsection{Magnetic field evolution during NS to BH collapse} 
\label{hair1}

During the merger of \NSs, a transient, super-massive, fast-rotating \NS\ is formed \citep[][]{2011ApJ...732L...6R}. After 10-100 milliseconds it collapses into a \BH. What are the possible observational effects of this collapse?

\begin{figure}[h!]
\includegraphics[width=\linewidth]{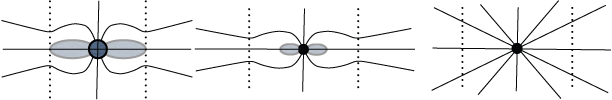}
\caption{Cartoon of the structure of magnetic fields around  a collapsing  rotating \NS. Initially, left panel, the \Bf\ is that of an isolated pulsar, with a set of field lines closing within the light cylinder (dashed vertical lines).  Immediately after the collapse, central panel, the structure is similar. The closed field lines are absorbed by the \BH, while the open field lines remain attached to the \BH; the system relaxes to the monopole structure (right panel), \cite{2011PhRvD..83l4035L}}
\label{NS-Collapse-picture}
\end{figure}

Importantly:   the magnetic energy contained within the collapsing  \NS\ \ms\ {\it is not} released an  \EM\ impulse during formation of a \BH, as would be predicted by the No-Hair theorem, Fig. \ref{NS-Collapse-picture}.
There is no EM pulse contemporaneous with the NS to BH collapse \citep{2011PhRvD..83l4035L}.    The collapse of a  NS into the  BH happens smoothly, without  natural formation of current sheets or other dissipative structures on the open field lines and, thus,  does not allow the \Bf\ to become disconnected from the  star and  escape to infinity.  As long as an  isolated Kerr   \BH\ can produce plasma and currents, it does not lose its  open \Bf\ lines, its magnetospheric structure evolved towards a split monopole  and the \BH\  spins down electromagnetically (the closed field lines get absorbed by the hole).

\begin{figure}[h!]
\includegraphics[width=\linewidth]{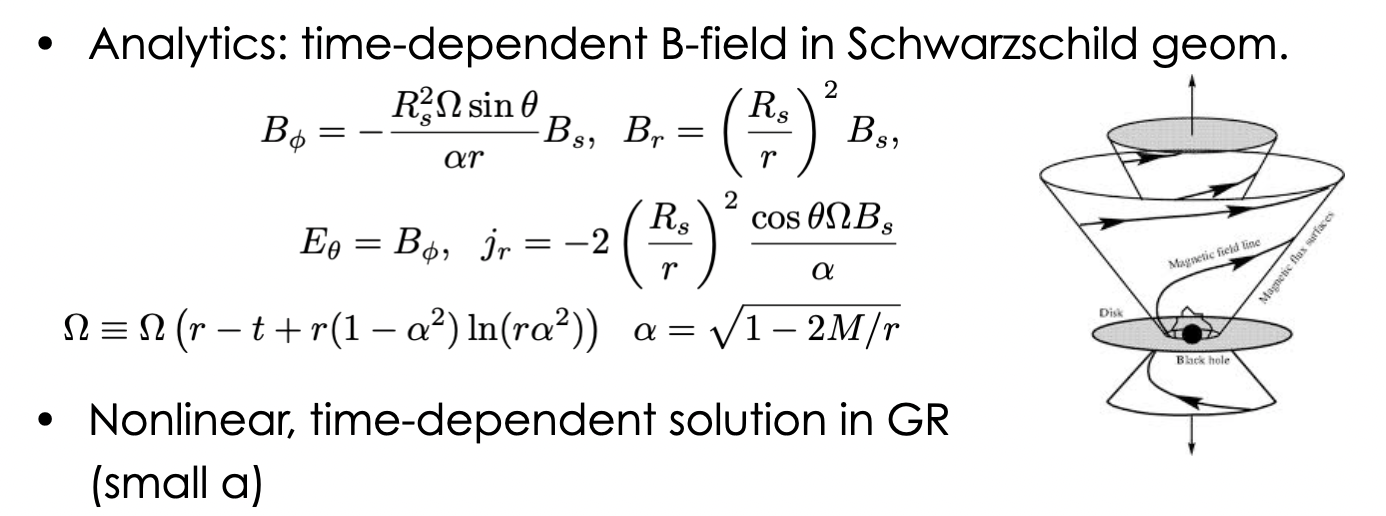}
\caption{Evolution of \Bf\ during NS $\to$  BH collapse  \citep{2011PhRvD..83l4035L}. Analytical solution for split-monopole \ms\ demonstrates that open field lines are not ``cut out'' during collapse.}
\label{Michel}
\end{figure}

To prove this point  \citep{2011PhRvD..83l4035L} generalized  the Michel's solution of monopolar \mss\ \citep{1973ApJ...180L.133M}  to  arbitrary $\Omega(r_{\rm fast}-t)$ in General relativity. The argument of $\Omega$ should be evaluated at the position of a radially propagating fast mode in the \Sch\  metric with  $dr_{\rm fast}/dt = \alpha^2$,  Fig. \ref{Michel}

It follows, that 
  the No-Hair theorem   is not formally  applicable for black holes formed from collapse of a
  rotating neutron star.  Rotating neutron stars can self-produce particles via vacuum
  breakdown forming a highly conducting plasma magnetosphere
  such that magnetic field lines are effectively ``frozen-in'' the star both before and during collapse.
  In the limit of no resistivity, this introduces a topological constraint which prohibits the magnetic field
  from sliding off the newly-formed event horizon. As a result, during
  collapse of a neutron star into a black hole, the latter conserves the number of
  magnetic flux tubes, hair. Hair can be counted (!)
\ba&&
N_B =  e \Phi_\infty /( \pi c \hbar) = { B_{NS}  e R_{NS} ^3 \Omega_{NS}}/({c^2 {\hbar}}) = 10^{41} {B_{NS} \over 10^{12} {\rm G}} \, {P_{NS}  \over 1 {\rm msec}} .
\nn &&
\Phi_\infty \approx  \pi  B_{NS} R_{NS}^3 \Omega_{NS} / c
\label{NB}
\ea
  $\Phi_\infty $ is the
  initial magnetic flux through the hemispheres of the progenitor and
  out to infinity, and $\Omega_{\rm NS} = 2\pi/P_{\rm NS}$ is pre-collapse spin of the \NS. 
  
\begin{figure}[h!]
  \includegraphics[width=0.99\linewidth]{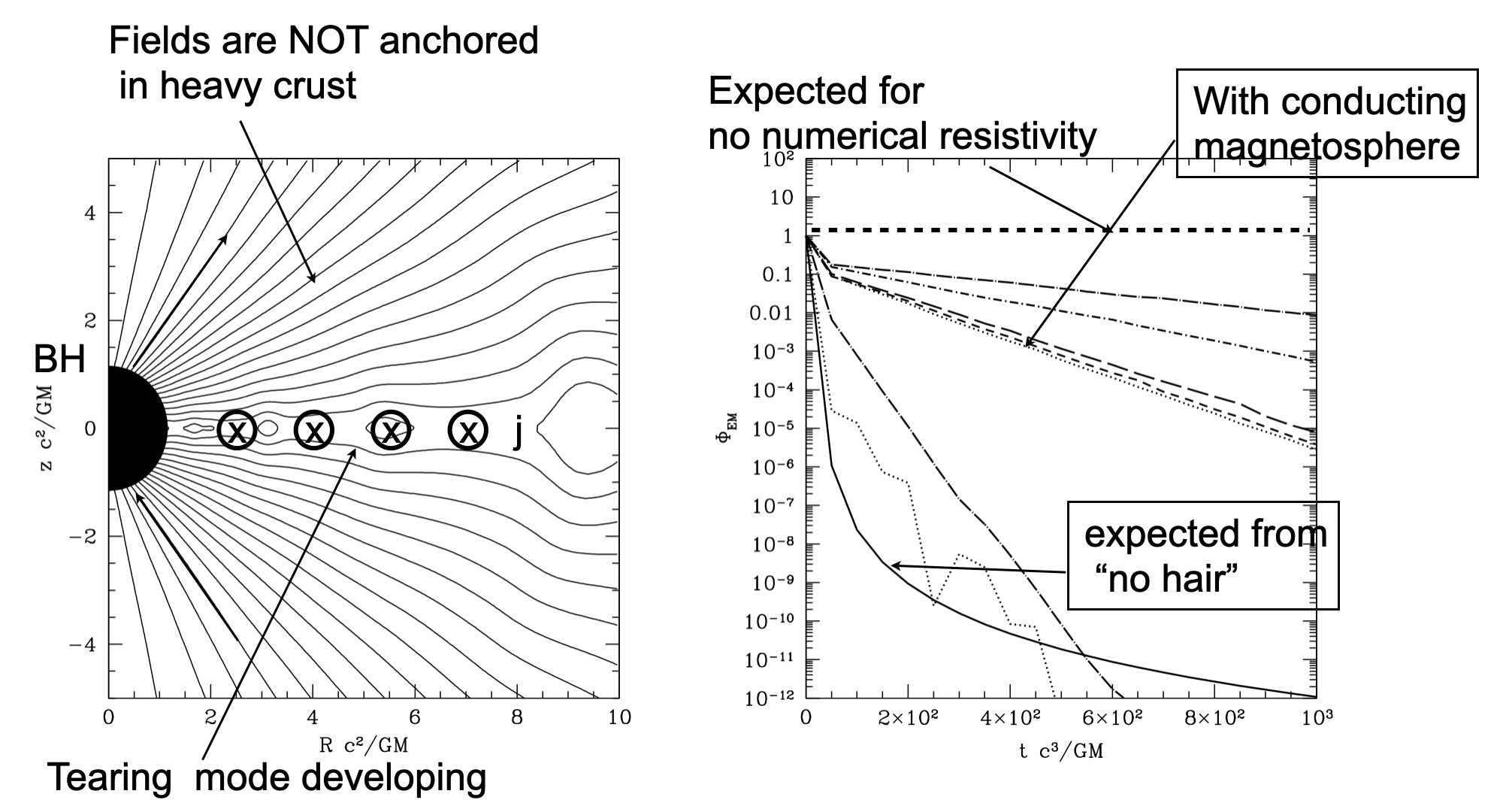}
   \includegraphics[width=0.9\linewidth]{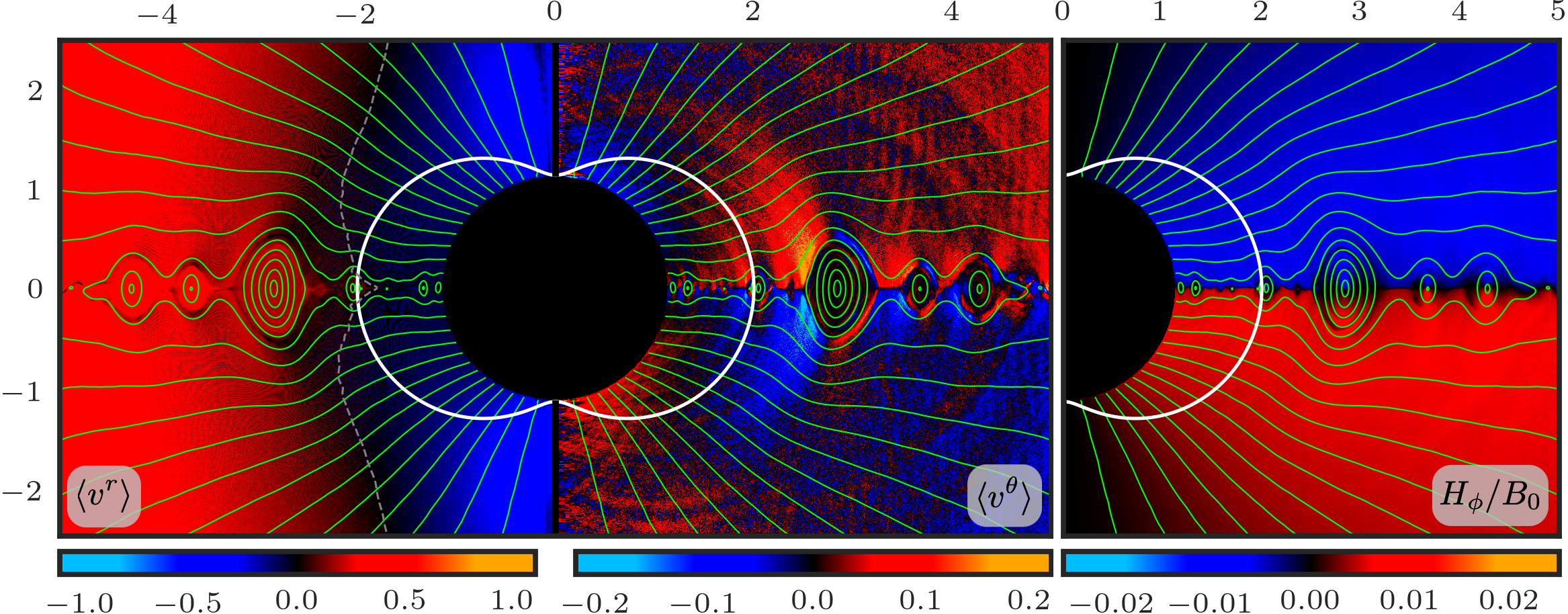}
  \caption{Magnetic hair of newly formed BH, numerical simulations. Top:  \cite{2011PhRvD..84h4019L}, annotated; bottom: \cite{2021PhRvL.127e5101B}, zoomed-in view of the equatorial reconnection current sheet.}
 \label{hair}
 \end{figure}

This theoretical result  was tested  by  \cite{2011PhRvD..84h4019L}  \citep[see also  latest simulations by][]{2021PhRvL.127e5101B,2023ApJ...956L..33M}, Fig. \ref{hair},  via 
  three-dimensional general relativistic plasma simulations of
  rotating black holes that start with a neutron star dipole magnetic field
  with no currents initially present outside the event horizon.  The
  black hole's magnetosphere  subsequently relaxes to the split monopole
  magnetic field geometry with self-generated currents outside the
  event horizon. The dissipation of the resulting equatorial current
  sheet leads to a slow loss of the anchored flux tubes, a process
  that balds the black hole on long resistive time scales rather than
  the short light-crossing time scales expected from the vacuum
  ``no-hair'' theorem. Eventually, reconnection events will lead to magnetic field release \citep[{\it Slowly balding black holes}, ][]{2011PhRvD..84h4019L,2023ApJ...956L..33M}, shutting down the electromagnetic BH engine forever.

The power produced by the resulting isolated BH (we stress - {\it isolated} - with \Bf\ not supported by accretion, but anchored in the  BH horizon by the ideal condition of the surrounding plasma) is
\be
L_h \sim \Phi_\infty^2 \Omega^2 /c =  6 \times 10^{44} \, {\rm erg\,  s} ^{-1}
\label{Lhair}
\ee
The  estimate of (\ref{Lhair}) is for \Bf\ of $10^{12} $ G. 
Power $L_h$  can be substantial, especially if \Bf\ is amplified by turbulence resulting from shear instabilities during the merger  \citep{2015ApJ...809...39G}. For example, for quantum \Bf\ and BH spin of 1 msec, the resulting ``hair-powered'' BH emission can be as high as $L_h \sim 3 \times 10^{48} $ erg s$^{-1}$.   Such process may contribute to the production of prompt emission in  short GRBs \citep{2013ApJ...768...63L}.

The post-collapse (magnetic)  ``hair-driven" stage is limited by how fast (or how slow) \Bf\ slides off the BH due to resistive effects in the equatorial current sheet. Numerical estimates  \citep{2011PhRvD..84h4019L,2023ApJ...956L..33M} indicate relatively rapid ``hair loss'' (still on time scales much longer than the dynamics time scale predicted by the No-Hair theorem. But resistive effects  in numerical simulations are typically larger than physically expected.

  \subsection{EM emission due to post-merger  BH's ringing in \Bf: the Gertsenshtein-Zeldovich effect}

 Gertsenshtein-Zeldovich (GZ) effect  is  the
coupling  of electromagnetic waves and  gravitational
waves in the presence of a strong magnetic
field \citep{gertsenshtein1962wave,1974JETP...38..652Z}.

Let a gravitational  wave propagate along $z$, while \Bf\ is in the $x-z$ plane. The \EM\ four-potential of the constant \Bf\  can be chosen as 
\be
A_\mu = \left\{0,-\frac{1}{2} y \cos (\theta ),\frac{1}{2} (x \cos (\theta )-z \sin (\theta
   )),\frac{1}{2} y \sin (\theta )\right\} B_0
   \ee
   (this form  of $A_\mu$ then fits the polarization of the gravitational waves).
   
   For x and + modes with amplitudes $h_{\rm x,+}$  the EM source evaluates to 
   \be
   S^\nu = \partial_\mu (\sqrt{-g} F^{\mu\nu}  )   = \{ 0,  h_{\rm x}  ,  -h_{\rm +}  , 0\} \times  i  \omega  B_0 \sin (\theta )e^{-i \omega  (t-z)}
\ee

Looking for the perturbed four-potential in the form
\be
A_\mu ^{(w)} = \{ 0, A_x (t) , A_y (t), 0\}   e^{-i (t \omega -\omega  z)}
\ee
(with slowly evolving amplitude),
we find that the amplitude of the EM will grow linearly with time according to 
\be
\partial_t A_\mu ^{(w)} = \{0, h_{\rm x}, -h_{\rm +}  ,0\} \times \frac{1}{2} B_0  \sin (\theta ) 
\ee

Corresponding EM Poynting flux, 
\be
{\bf F} =  \{ h_{\rm +}  ^2, h_{\rm +}  h_{\rm x},0\}  \times   \frac{B_0^2 t^2 \omega ^2 \sin ^2(\theta )}{32 \pi },
\ee
has  the following properties
\begin{itemize}
\item
Gravitational waves propagating along the external \Bf\ do not decay
\item   Higher frequencies decay faster, $\propto \om^2$.
\item
EM emission correlates with GW  polarization (and the direction of \Bf). 
\end{itemize}

The power of the  magnetic  ring-down emission   can be estimated as 
 \be
L_B  \approx  h_{\rm +,x} ^2  (B_0 R_g)^2 c  = 5 \times 10^{46 } \,  {\rm erg \, s}^{-1}  \times  (h/0.1)^2 (B/B_Q)^2 (M/M_\odot)^2 
\label{ringdonw}
 \ee
 for $\om \sim c/R_g$. It can be mildly powerful.
 Qualitatively, 
 \be
 \frac{L_B} {L_{GW,r} }  \approx \frac {B_0 ^ 2 R_g^3} { M c^2} \sim 3  \times 10^{-12} (B/B_Q)^2 (M/M_\odot)^2
 \ee
 where  $L_{GW,r}$ is the power coming out in ringdown gravitational waves (both scale as $ h_{\rm +,x} ^2 $), and $B_Q$ is quantum fields. 
    Thus, the overall power magnetic of   ring-dow  is  small, unlikely to affect the LIGO/LISA waveforms.

 \section{Magnetospheric interaction in BH-NS mergers}
\label{BH} 
The \cite{1977MNRAS.179..433B} paradigm outlines how the energy of a rotating \BH\ can be extracted to generate jets and the observed broadband emission. There is another possibility: Schwarzschild black holes as unipolar inductors \citep{2011PhRvD..83f4001L,2014PhRvD..89j4030M}. 

\begin{figure}[h!]
  \includegraphics[width=0.32\linewidth]{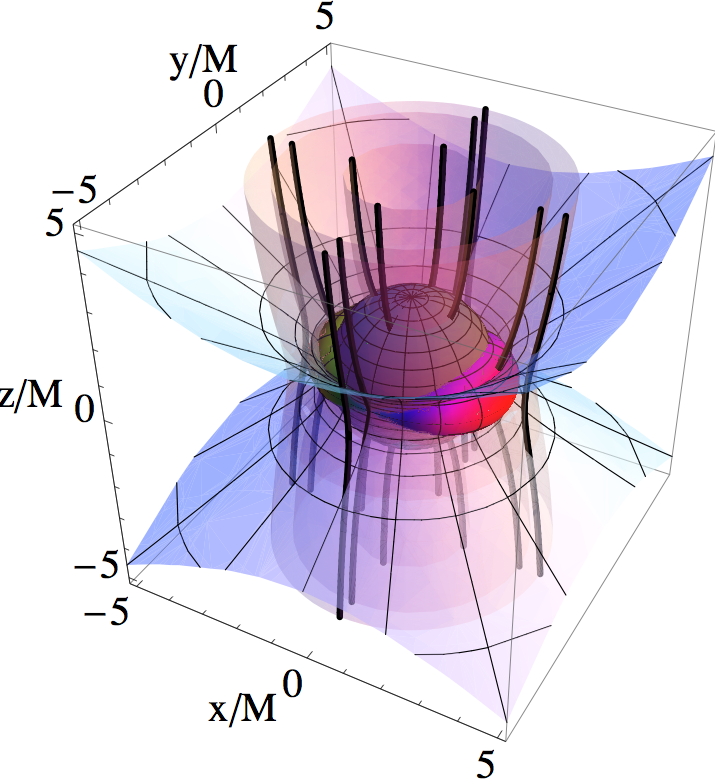}
   \includegraphics[width=0.32\linewidth]{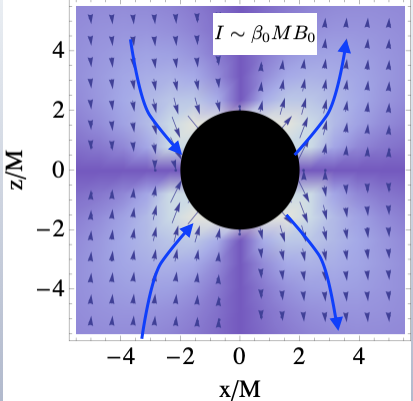}
   \includegraphics[width=0.32\linewidth]{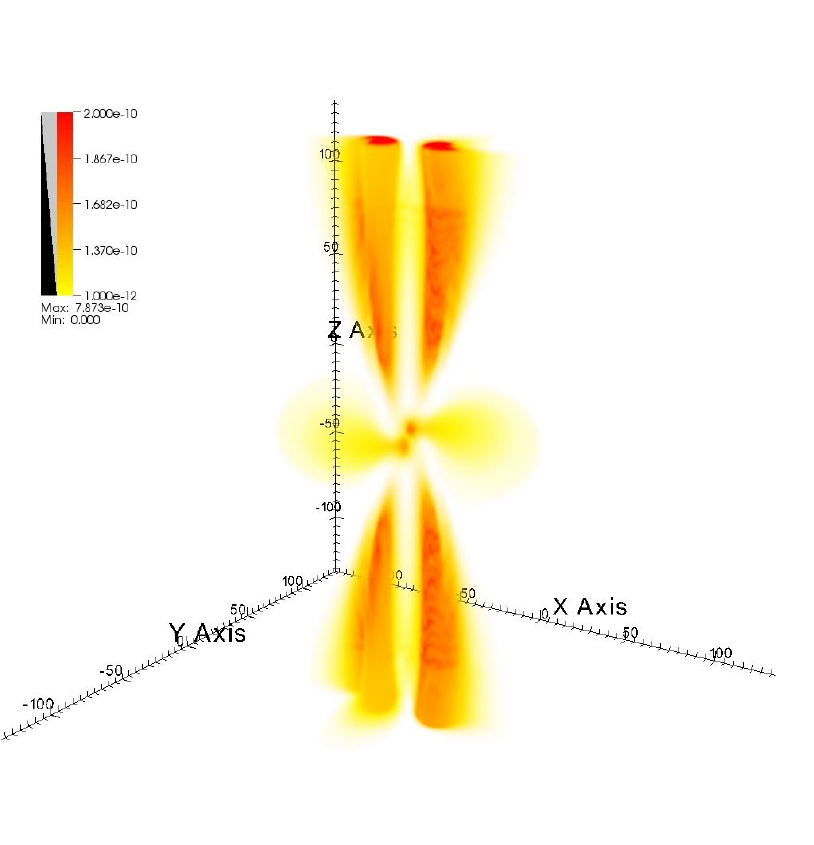}
  \caption{Left and center panels: 3D view of the magnetosphere of a \BH\ moving linearly through magnetic field \citep[after][]{2011PhRvD..83f4001L}. A BH generate {\it  quadruple} current flow.  Right panel: GR-MHD simulations  of merging black holes in external \Bf\ by \cite{2010Sci...329..927P}. (We point out that the corresponding PIC simulations have not been done so far.)}
 \label{bigpicture}
 \end{figure}

Two related processes are expected to be at play. First, the motion of a Schwarzschild black hole  through a constant magnetic field  $B_0$ in vacuum induces a component of the electric field along  the magnetic field, $\E \cdot \B \neq 0$. 
 Another breakdown of the ideal condition is   the  creation of regions where electric field becomes larger than \Bf, Fig. \ref{bigpicture}.
 
Regions of $ \E \cdot \B \neq 0$ and $E>B$  will lead to dissipation via radiative effects and vacuum breakdown. To bring the system ``back to normal'', an electric charge density will be produced,  of the order of $\rho _{\rm ind}= B_0 \beta_0 /( 2 \pi e R_G)$, where $R_G =  2 G M/c^2$ is the Schwarzschild radius and $M$ is mass of the \BH; the charge density $\rho _{\rm ind}$ is similar to the Goldreich-Julian density. 

As a result, the magnetospheres of moving {\BH}s resemble in many respects the magnetospheres of rotationally-powered pulsars, with pair formation fronts and outer gaps, where the sign of the induced charge changes.
The black hole generates bipolar electromagnetic jets each consisting of two counter-aligned current flows (four current flows total), each carrying an 
 electric current of the order $I \approx e  B_0 R_G \beta_0$, Fig. \ref{bigpicture}. The electromagnetic power of the jets is $L \approx (G M)^2 B_0^2 \beta_0^2/c^3$ \cite[][]{2011PhRvD..83f4001L,2023ApJ...956L..33M}.

Since the resulting electrodynamics is in many respects similar to pulsars, 
 merging black holes may generate coherent radio and high-energy 
 emission beamed approximately along the orbital normal \cite[][]{2011PhRvD..83f4001L,2023ApJ...956L..33M}.

For \BH\ moving linearly through \Bf, the  relations  (\ref{L}) gives (assuming orbital motion is dominated by the BH)
\ba &&
L_{BH-NS} = \frac{ ( G M_{BH})^3 B_{NS}^2 }{\pi c^5 R_{NS} } \left( \frac{R_{NS}} {r} \right) ^7=
 \left\{
 \begin{array}{c}
 3 \times 10^{46} \, {\rm erg\,s}^{-1}\,  m_1^3 \,  \left( \frac{R_{NS}} {r} \right) ^7
 \\
 5 \times 10^{35} \, {\rm erg\,s}^{-1}\,  m_1^{-1/2} \,   \left( \frac{-t}{ {\rm sec} }  \right)^{-7/4},
 \end{array}
 \right.
 \nn &&
 m_1 = \log_{10}  \frac{M_{BH}}{ M_\odot} \geq 1
 \ea

\section{Observation strategy}

Realistically, it's a huge  challenge to detect precursor emission to LVK sources. Yet the possible information to be gained - detailed properties of the merging objects  - would be highly valuable to constraint the physical properties of the compacts, and by extension the  larger astronomical  picture of evolution of stellar remnants. As we discuss next, some (less likely) channels are already in places, some will require dedicated observing strategy. Yet the corresponding costs - {\eg} numerically targeting large-view radio phased array telescopes  - are comparatively mild.

\subsection{Best chance: low frequency  radio}

The best chance of seeing precursor emission is the production of coherent pulsar-like  radio  emission \citep{2001MNRAS.322..695H,2019MNRAS.483.2766L,2023MNRAS.519.3923C}. For example, if peak power is emitted in radio with efficiency $\eta_R = 10^{-3} \eta_{R,-3} $, the expected observed flux may be in the Jansky range:
\be
F_{\nu, peak} = 0.5 \, {\rm Jy}  \eta_{R,-3}  \nu_{9} ^{-1} d_{200}^{-2}
\ee
where $\nu = 10^ 9 \nu_{9}  $ Hz is the observed frequency and $d_{200} =d/(200) \, {\rm Mpc}$ is distance to the source. 
Thus, in the best case scenarios the radio power can be substantial. 
 
 In addition to    LVK early warnings alerts \citep{2020ApJ...905L..25S}, $\sim$ one minute advanced notice, 
 radio pulses  have another advantage: they will experience dispersive delay of
 \be
 \Delta t = 14\, {\rm sec} \nu_9^{-2}  d_{200}
 \ee
 \citep[see also][]{2010Ap&SS.330...13P}.
 At low frequencies, $\leq  100 $ MHz,  the delay can be nearly half an hour!  \citep[][discussed MWA detectability of LVK  precursors]{2023PASA...40...50T}. The expected localization  of early alerts is of the order of 1000 square degrees. Such relatively wide field requires particular observing mode, similar to the digital beamforming  of $\sim 30$ square degrees used to localize a single FRB

The radio emission is expected to be modulated on the spin period(s) of the merging {\NS}s, as well as orbital-spin beats, \S \ref{2to1}. Dispersion measure will be similar to FRBs, in the hundreds. \cite[Mergers were the leading model of FRBs before  the Repeater was discovered][.]{2016Natur.531..202S}  Potentially problematic  for detection of periodicity is the orbital evolution during last orbits (when emission is most powerful), as the beat frequencies evolve with time. This introduces complication for the de-dispersion process.

\subsection{High energy: beaming needed}

\begin{figure}[h!]
  \centering
   \includegraphics[width=0.8\linewidth]{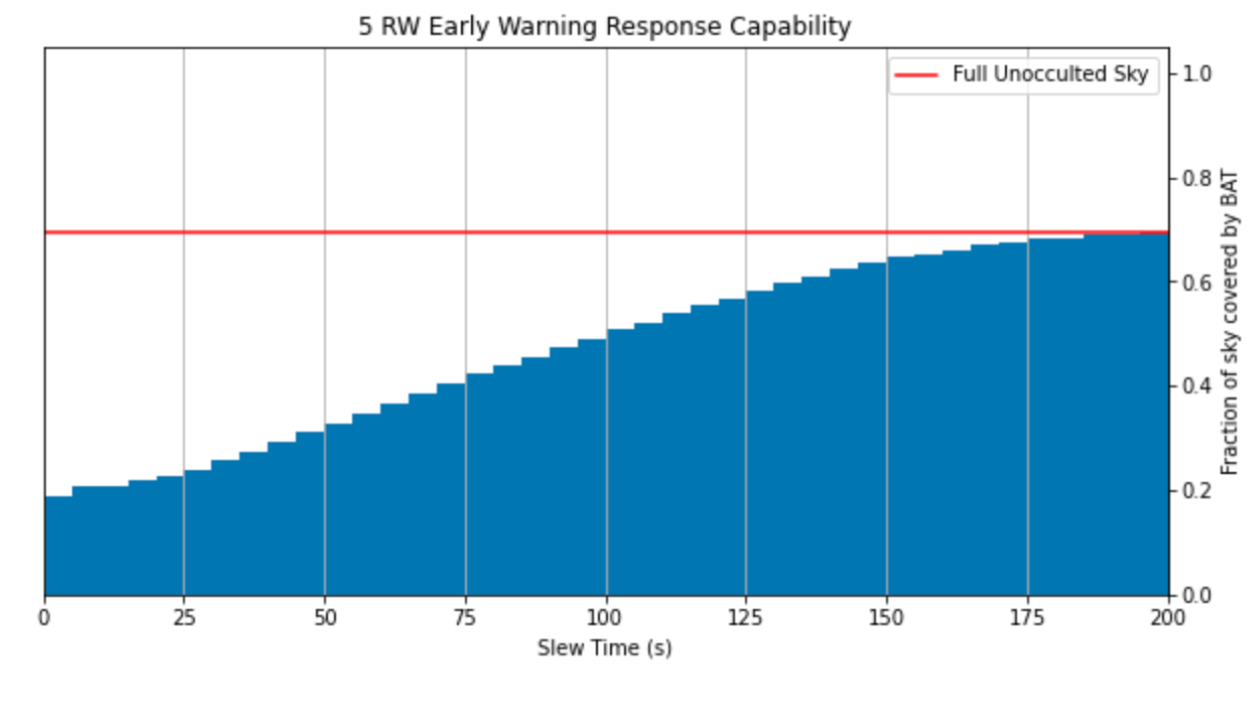}
   \caption{Fraction of sky BAT can cover as a function of latency \citep{2020ApJ...900...35T}.  If alert comes out at T0-60 seconds, Swift can get to the right position before merger in $\sim 40\%$ of cases.}
 \label{earlywarning-swift}
 \end{figure}

Isotropic powers (\ref{L1},\ref{Lpeak})  are not sufficient to be detected by high energy all-sky monitors satellites.  Unknown location does not allow use of more sensitive  targeted instruments. In case of Swift, The SWIFTGUANO
\citep{2020ApJ...900...35T} pipeline may achive sufficiently fast slewing, starting as early as  8 seconds from the trigger,  Fig. \ref{earlywarning-swift}

Beaming can be  another possibly positive  effect.  Emission produced in $\E \cdot \B \neq 0$ gaps will be beamed along the local \Bf.  Beamed emission reduces chances of been seen, but increases the flux. Both NS-NS and NS-BH mergers   are  expected to produce beamed emission.

\subsection{Optical}

Survey  optical telescopes typically have fields of ten(s) square degrees. They are unlikely to be just looking at the right direction. LIGO early alerts open a possibility.

Optical flash of $\sim 10^{43}$ erg s$^{-1}$ coming from 200 Mpc would correspond to apparent magnitude of $m \sim 15$, but only for a millisecond
\citep[see a closely related discussion of FRB optical detectability][] {2016ApJ...824L..18L}. Fast read-out is the key then.
If the readout time is ten seconds, a millisecond flash
will give a fluence $\approx 10^4$ times smaller resulting in 
an image 10 magnitudes fainter, $m\sim 25$. This is  somewhat below Zwicky Transient Facility   \citep[ZTF,][]{2019PASP..131g8001G,2019PASP..131a8002B}.

ASAS-SN \citep{2016arXiv160400396H},
EVRYSCOPE \citep{2015PASP..127..234L}, DECam \citep{2018ApJS..234...39S}, ATLAS \citep{2018PASP..130f4505T},  Pan-STARRS \citep{2002SPIE.4836..154K}  seem to be below the required  threshold. (DECam is the most sensitive of them all, but has  the field of view of  "only" 3 deg. sq.). The  Argus Array \citep[''Evryscope on steroids''][]{2022PASP..134c5003L} is highly promising.

The forthcoming Large Synoptic Survey Telescope
\citep[LSST;][]{2008arXiv0805.2366I} is expected to fare  better.  It will also
have large field of view, almost 10 square degrees,
and will be able to reach magnitude
$m=25$ within two exposures of $\sim 15$~s each.

 The  analysis of LSST  data would be  problematic, though: the image will be only on one plate, close to the limit of LSST sensitivity. But 
 there are a lot of optical transients ({\eg} asteroids passing by), so that single-image sources are typically discarded.

An important fast readout procedure,  the 
Readout While Exposing mode \citep[RWE,][]{2009AJ....138..568B},  is been tested  on wide-field instruments, like Megacam \citep{2009AJ....138..568B}  and ZTF (Andreoni et al., in prep).  It is expected to reach a few milliseconds readout rate (at the  expense of sacrificing  one dimension in the image). Some prelimnary  testing is going on (Phinney {\etal}, priv. comm.).  It is not planned to be implemented at LSST, though.

Another possibly interesting optical wide-field *and* fast read-out  observatory is Tomo-e Gozen \citep{2018SPIE10702E..0JS,2020PASJ...72....3R,2024MNRAS.527..334O} telescope, with 10-500ms time resolution.
 
\section{Supermassive BH-BH mergers and LISA precursors}

Merger of supermassive BHs - target of future  LISA observations - may also produce precursor emission \citep{2010Sci...329..927P,2011PhRvD..83f4001L}. The difference from the NS-BH merger is that the \Bf\ needs to be supplied by the external accretion disk. In addition, accretion contribution - as opposed to the purely electromagnetic discussed here - may be considerable \citep{2005ApJ...622L..93M,2009MNRAS.393.1423C,2017ApJ...835..165B}. As the EM power is concerned,  in Eq. (\ref{L})  the  \Bf\  is then parametrized by the properties of the accretion disk ({\eg} mass accretion rate, $\alpha$-viscosity,  and location of the inner edge), while the velocity if the Keplerian velocity of the merging BHs. 

The power of unipolar induction, for a given external \Bf, remains mild. For example,  comparing the power of the  \Sch\ \BH\ as unipolar inductor,  with the Blandford-Znajek power in a given \Bf, $L_{BZ} \sim a^2 B^2 M^2$ ($a$ is the \BH\ spin parameter), 
 gives
 \be
 {L_{EM} \over L_{BZ}} \approx \frac{\beta_K^2}{ a^2}
 \ee
 ($\beta _ K$ is dimensionless Keplerian velocity.  For fast rotating {\BH}s, $a \sim 1$,  the power of the  unipolar inductor is subdominant until right before the  merger.
 
 In addition, merging supermassive BHs may produce plerionic-type emission \cite{2011PhRvD..83f4001L}. Most of the EM  power  will leave the {\BH}s' region in a form of relativistic highly magnetized wind. Even though the  instantaneous  power  is typically much smaller than the Eddington power corresponding to masses $M$, 
 the total released energy can be substantial. 
 
 An important, and uncertain, parameter is the decoupling radius, when the gravitational radiation acts faster  on the BH separation than the accretion time scale of the inner disk. 
 Overall, one can expect $\sim 10^{47} \, {\rm erg} \, m_6^{9/5} $ ergs deposited \citep[see ][for estimates]{2011PhRvD..83f4001L}.

The Nanograv connection. 
Evidence for a low-frequency stochastic gravitational-wave background has recently been reported by {\it Nanograv} based on analyses of pulsar timing array data \citep{2023ApJ...951L...8A}. The most likely source of such a background is a population of supermassive black hole binaries, the loudest of which may be individually detected in these data sets
\citep{2023ApJ...951L..50A}. Two candidate sources showed marginal/weak evidence of detection.

Objectively, in radio, observations of \EM\ from supermassive binary black holes merger is a difficult task due to long periods (years) and poor localization even if a particular single source can be identified. In radio, it is hard to control for all the
systematics over the years. 

LSST may offer a better coverage: it is expected that after 5 years of LSST observations, tens of true binaries will be detectable, as periodically variable AGNe \citep{2019BAAS...51c.490K,2019MNRAS.485.1579K}. Given the period from Nanograv and LSST coverage, one may expect/hope to identify \EM\ signatures of merging \BH.

\section{In conclusion}

Realistically, detecting precursor emission to LVK sources is highly challenging, yet the corresponding costs  in X-ray and radio -  numerically targeting large-view radio phased array telescopes, and   SWIFTGUANO - are comparatively mild. Optical dectections are the most challenging.

 \section{Disclaimer and Acknowledgment}
These notes are not intended as a review of the field, or  formal publication. Future discussions/collaborations are  encouraged. 

 I would like to thank  Igor Andreoni for most enlightening discussions. I also thank  Shri Kulkarni, Elias Most,   Sterl Phinney  and  Aaron Tohuvavohu for comments.

\bibliographystyle{apsrev}

 \bibliography{/Users/maxim/Dropbox/Research/BibTex,/Users/maxim/Dropbox/Research/BibTex-1}

\end{document}